# The temperature and chronology of heavy-element nucleosynthesis in low-mass stars

P. Neyskens†, S. Van Eck†, A. Jorissen†, S. Goriely†, L. Siess† & B. Plez‡

†Institut d'Astronomie et d'Astrophysique, Université libre de Bruxelles (ULB), CP 226, 1050 Bruxelles, Belgium
‡LUPM, Université Montpellier 2, CNRS, F-34095 Montpellier, France

**Roughly half of the elements heavier than iron in the Universe are believed to be synthesized in the late evolutionary stages of stars with masses between 0.8 and 8 solar masses. Deep inside the star, nuclei (mainly iron) capture neutrons and progressively build up ('s-process'[1,2]) heavier elements which are recurrently 'dredged-up' towards the stellar surface. Two neutron sources, activated at distinct temperatures, have been proposed: $^{13}C(\alpha, n)^{16}O$ or $^{22}Ne(\alpha, n)^{25}Mg$. To explain measured stellar abundances[1–7], stellar evolution models invoking the $^{13}C(\alpha, n)^{16}O$ neutron source which operates at temperatures of ~ $10^8$K are favored. However, isotopic ratios in primitive meteorites, reflecting nucleosynthesis in previous generations of stars, point at higher temperatures (above $3 \times 10^8$K)[1], requiring at least a late activation of $^{22}Ne(\alpha, n)^{25}Mg$. Here we report on a determination of the s-process temperature directly in stars, using accurate zirconium and niobium abundances, independently of stellar evolution models. The derived temperature clearly supports $^{13}C(\alpha,n)^{16}O$ as the s-process neutron source. The radioactive pair $^{93}Zr$ – $^{93}Nb$ used to estimate the s-process temperature also provides, together with the $^{99}Tc$ – $^{99}Ru$ pair, chronometric information on the time elapsed since the start of the s-process (1–3 Myr).**

We obtained high-resolution spectra of 17 S stars and of 6 M stars with the high-resolution HERMES spectrograph[8] ($\lambda/\Delta\lambda$ = 80 000, Extended Data Table 1 and Extended Data Fig. 1). S stars are s-process-enriched red giants with effective temperatures in the range 3000 to 4000 K, while M stars are similar giant stars but showing no s-process enhancement. Stellar parameters (effective temperature, surface gravity, carbon-to-oxygen ratio, s-process enhancement, and metallicity [Fe/H] = $\log_{10}(N(Fe)/N(H))_{star}$ - $\log_{10}(N(Fe)/N(H))_{\odot}$, where $N(A)$ is the number density of element A and $\odot$ denotes the Sun), are determined by comparing observational data with predicted spectra and photometric colors[9], computed from a grid of dedicated model atmospheres[10, 11]. Fe, Zr, Nb and Tc abundances are derived, as well as the corresponding errors arising from estimated uncertainties on the stellar parameters (Extended Data Tables 2 – 5 and Extended Data Fig. 2).

Actually, S stars come in two varieties according to the presence or absence of Tc, a chemical element with no stable isotope. *Extrinsic* S stars lack Tc and are all binaries[12]. The atmosphere of these giant stars contains s-process material transferred from a companion that has completed its path through the asymptotic giant branch phase, and therefore reflects the entire s-process production history. We use the $N(Nb)/N(Zr)$ ratio in these extrinsic S stars to derive the s-process temperature.

Mono-isotopic Nb can only be produced by β-decay of $^{93}Zr$. In extrinsic S stars, the time elapsed since the end of the mass transfer is much longer than the $^{93}Zr$ half-life ($\tau_{1/2}$ = 1.53 Myr). Consequently, the $N(Zr)/N(Nb)$ abundance measured today in the extrinsic S star is equal to the $N(Zr)/N(^{93}Zr)$ abundance at the end of the s-process in the companion, provided the s-process contribution dominates over the initial heavy element abundances, which is the case in the most enriched extrinsic S stars.

As we show in Section 5 of the Methods (Eqs. 12 and 15), the $N(Zr)/N(^{93}Zr)$ ratio is directly related to the neutron-capture cross sections of the various Zr isotopes, under the approximation of local equilibrium for the s-process flow. Since these neutron-capture cross sections depend on temperature, so does the $N(Zr)/N(^{93}Zr)$ ratio at the s-process site. *$N(Zr)/N(Nb)$ is thus a measure of the s-process temperature*.

Fig. 1 compares the $N(Zr)/N(Nb)$ ratio, as predicted from the temperature-dependent neutron-capture cross sections (squares), with those measured in the most enriched extrinsic S stars (dashed line). Fig. 1 also indicates the temperature ranges for the operation of the $^{13}C(\alpha, n)^{16}O$ and $^{22}Ne(\alpha, n)^{25}Mg$ neutron sources. In order to reconcile, within 1-$\sigma$ uncertainty, the predicted $N(Zr)/N(Nb)$ ratio with the one measured in extrinsic S stars, the s-process operation has to take place below ~ $2.5 \times 10^8$K. This is compatible with the operation of the $^{13}C(\alpha, n)^{16}O$ reaction as the neutron source but disfavors the $^{22}Ne(\alpha, n)^{25}Mg$ reaction. This confirms theoretical expectations[1–7] that protons from the outer convective envelope diffuse in the C-rich region left after the development of the thermal pulse, which is a recurrent convective instability occurring in the He-burning shell of asymptotic giant branch stars. This mixing process still has a debated origin but is necessary to activate the chain of reactions $^{12}C(p,\gamma)^{13}N(\beta^+)^{13}C(\alpha, n)^{16}O$ responsible for the large neutron irradiation required to synthesize s-process elements[3].

The present determination of the s-process temperature relies on a single assumption, namely the equilibrium approximation along the Zr isotopic chain which is known to be valid locally[1]. For this reason, the uncertainties of the method are mainly those of the derived abundances and of the Zr experimental cross section[13, 14], i.e about 5% for the stable Zr isotopes and 11% for $^{93}Zr$. Reducing the $^{93}Zr$ error would constrain even more the s-process operation temperature.

Our result offers a long-sought confirmation of the expected – albeit still uncertain in its detailed modeling – operation for the s-process inside stars on the asymptotic giant branch phase. However it contradicts previously-reported higher s-process temperatures that were determined from isotopic abundance patterns in the solar system[1, 2], essentially derived from primitive carbonaceous chondrite meteorites of type C1. In contrast to the $N(Zr)/N(Nb)$ ratio in extrinsic S stars which probes a single s-process site (i.e., a single star), the solar-system isotopic abundance ratios represent a mix of diverse nucleosynthetic events that took place in stars of different masses and metallicities, integrated over the Galactic evolution, until the formation of the protosolar cloud. Therefore, their interpretation is less straightforward. They lead to high s-process branching temperatures that are difficult to reconcile with stellar nucleosynthesis in low-mass stars, where stellar models predict the s-process to occur at rather low temperatures during the interpulse phase. Instead, previous studies[1] inferring high s-process temperatures had to invoke a brief (a few years) neutron burst through $^{22}Ne(\alpha, n)^{25}Mg$ in a late hot thermal pulse, just sufficient to imprint a specific isotopic signature at the s-process branchings but not affecting the overall abundance distribution. The present method is insensitive to such a late and unrepresentative signature of the s-process. It rather probes the interpulse temperature, where the bulk of the s-process nucleosynthesis takes place.

Abundances in S stars do not only act as a *thermometer*, but also as a *chronometer*, based on the $^{93}Zr - ^{93}Nb$ and $^{99}Tc - ^{99}Ru$ decays in the second variety of S stars, the *intrinsic* S stars.

These stars are in the asymptotic giant branch phase of their life and are enriched in Tc[12,15] (Extended Data Fig. 3). $^{99}$Tc is the only Tc isotope produced by the s-process, and because of its rather short half-life ($\tau_{1/2}$ = 0.21 Myr), its detection is a signature of an on-going s-process nucleosynthesis. However the exploitation of Tc abundances has proven difficult[16,17], mainly because reliable abundances are challenging to obtain in cool stars. The Nb and Tc abundances derived for intrinsic S stars, as explained in Section 2 of the Methods, yield timescales with meaningful error bars (Table 1). Indeed, obtaining ages with sufficiently small error bars is always challenging when dealing with nucleo-cosmochronometers.

These timescales are counted from the first thermal pulse to the present time. This is shown in Fig. 2 and Extended Data Fig. 4, where observed $N(Tc)/N(Zr)$ and $N(Nb)/N(Zr)$ abundance ratios (dashed line with ±1$\sigma$ error bars represented as solid lines) are compared with model predictions for stars on the asymptotic giant branch phase accounting for the production and decay of $^{99}$Tc – $^{99}$Ru and $^{93}$Zr – $^{93}$Nb (panels A and B, respectively, of Fig. 2). In each model, for each time, a weight is assigned, separately for $N(Tc)/N(Zr)$ and $N(Nb)/N(Zr)$, depending on the difference between the observed and predicted abundance ratios (equation 3 and Section 4 of the Methods). It ranges from 0 (infinite difference) to 1 (no difference). The best simultaneous agreement between measured and predicted abundances of Tc *and* Nb is obtained when the product of the weights is maximal; this provides the asymptotic giant branch evolutionary timescale (panel C of Fig. 2), along with the corresponding model parameters (Table 1 and Extended Data Table 6).

An independent validation of the derived timescales comes from the infrared excess. As the star evolves along the asymptotic giant branch phase, mass loss, pulsations and low surface temperatures allow dust to condense, manifesting itself as infrared excesses[18]. Therefore, we expect – and actually see (Fig. 3 and Table 1) – a correlation between the asymptotic giant branch timescales of intrinsic S stars, counted from the first thermal pulse, and the infrared excesses $R = F(12~\mu m)/F(2.2~\mu m)$ probing the presence of circumstellar material emitting at 12 $\mu$m with respect to the stellar continuum at 2.2 $\mu$m. Although more data points are highly desirable, such a time-calibration of infrared excesses is precious because infrared data are more easily obtained for these cool objects than accurate abundances of radioactive chronometers. In the future, this relation could be used as an efficient proxy for measuring asymptotic giant branch timescales for stars covering a limited mass range, as it appears to be the case (Table 1) for intrinsic S stars.

**References**

1. Käppeler F., The origin of the heavy elements: the s process, *Prog. Part. Nucl. Phys.* **43**, 419 – 483 (1999).

2. Käppeler F., Gallino R., Bisterzo S., Aoki W., The s process: Nuclear physics, stellar models, and observations, *Reviews of Modern Physics* **83**, 157–193 (2011).

3. Straniero O., *et al.*, Radiative C-13 burning in asymptotic giant branch stars and s-processing, *Astrophys. J. L.* **440**, L85 – L87 (1995).

4. Goriely S., Mowlavi N., Neutron-capture nucleosynthesis in AGB stars, *Astron. Astrophys.* **362**, 599 – 614 (2000).

5. Van Eck S., Goriely S., Jorissen A., Plez B., Discovery of three lead-rich stars, *Nature* **412**, 793 – 795 (2001).

**Acknowledgements**

P.N. acknowledges the support of a FRIA (FNRS) fellowship. S.G., L.S. and S.V.E. are F.R.S.-FNRS Research Associates. B.P. is supported in part by the CNRS Programme

National de Physique Stellaire. This work has been partly funded by an *Action de recherche concertée* (ARC) from the *Direction générale de l'Enseignement non obligatoire et de la Recherche scientifique, Communauté française de Belgique*. Based on observations obtained with the HERMES spectrograph, supported by the Fund for Scientific Research of Flanders (FWO), the Research Council of K.U.Leuven, the *Fonds National de la Recherche Scientifique* (F.R.S.-FNRS), Belgium, the Royal Observatory of Belgium, the *Observatoire de Genève*, Switzerland and the Thüringer Landessternwarte Tautenburg, Germany.

**Author Contributions**

S.V.E. and A.J. initiated the project. S.V.E. was supervisor of the Ph.D. thesis of P.N., who computed the stellar atmospheric parameters, the abundances, and the grid of S-star model atmospheres with help from B.P. The stellar models were computed by L.S. and the nucleosynthesis predictions by S.G. The text was written by S.V.E. and A.J., and edited by the other authors.

**Author Information**

Reprints and permissions information is available at www.nature.com/reprints. The authors declare no competing financial interests.
Correspondence and requests for material should be adressed to S.V.E. (svaneck@astro.ulb.ac.be).

**Table**

**Table 1:** Derived ages and masses for our sample of intrinsic S stars.

| Star Name | Mass loss | $t_{TP}$ (Myr) | $t_{TP,min}$ (Myr) | $t_{TP,max}$ (Myr) | $t_S$ (Myr) | $M_{mod}$ ($M_\odot$) | $\log(L/L_\odot)$ | C/O | [Fe/H] |
|---|---|---|---|---|---|---|---|---|---|
| $o^1$ Ori | VW | 1.3 | 0.7 | 2.6 | 0.6 | 2 | 3.45 | 0.50 | -0.45 |
|  | S | 1.8 | 0.7 | 3.9 | 0.6 | 3 |  |  |  |
| AA Cam | VW | 1.4 | 0.9 | 2.3 | 0.6 | 3 | 3.91 | 0.50 | -0.40 |
|  | S | 1.6 | 0.7 | 2.7 | 0.4 | 3 |  |  |  |
| KR CMa | VW | 1.6 | 0.7 | 2.6 | 0.8 | 3 | - | 0.50 | -0.34 |
|  | S | 1.8 | 0.7 | 3.9 | 0.6 | 3 |  |  |  |
| CSS 454 | VW | 1.7 | 1.1 | 2.6 | 1.0 | 2 | - | 0.50 | -0.40 |
|  | S | 1.7 | 1.0 | 3.9 | 1.0 | 2 |  |  |  |
| HIP 103476 | VW | 2.2 | 1.2 | 3.2 | 1.4 | 3 | 3.59 | 0.50 | -0.01 |
|  | S | 2.5 | 1.5 | 4.0 | 1.3 | 3 |  |  |  |
| AD Cyg | VW | 2.7 | 2.1 | 3.9 | 1.9 | 3 | - | 0.97 | -0.05 |
|  | S | 3.0 | 2.2 | 4.0 | 1.8 | 3 |  |  |  |
| NQ Pup | VW | 2.8 | 2.1 | 4.2 | 2.0 | 3 | 2.95 | 0.50 | -0.31 |
|  | S | 3.1 | 2.3 | 4.0 | 1.9 | 3 |  |  |  |
| HR Peg | VW | 3.2 | 2.4 | 4.8 | 2.4 | 3 | 3.39 | 0.75 | 0.00 |
|  | S | 3.0 | 2.4 | 3.7 | 1.8 | 3 |  |  |  |

Mass-loss rates are known to influence the stellar evolution and the minimum initial mass needed for producing intrinsic S stars. Two grids of models were thus computed, corresponding to either VW[19] or S[20] mass-loss prescriptions. The impact on the derived ages is rather limited. $t_{TP}$ (resp., $t_S$) stands for the ages counted from the first thermal pulse (resp., from the time at which s-process elements enrich the stellar surface). $M_{mod}$ stands for the stellar mass of the model yielding the best fit to the abundances. $L$ is the luminosity of the star[21]. The C/O ($\equiv N(C)/N(O)$) and [Fe/H] values providing the best agreement with observations are also listed.

**Figure legends**

Figure 1: **The $^{93}$Zr – $^{93}$Nb pair as an s-process thermometer.**

We compare $N(Zr)/N(Nb)$, i.e. $N(Zr)/N(^{93}Zr)$, deduced from experimental Maxwellian-averaged cross sections at given temperatures (full squares) with the value measured in the most enriched S stars (dashed line), with standard error (hatched zone) on the orthogonal distance regression (Eq. 14 of the Extended Data). The temperature above which the $^{13}$C and $^{22}$Ne neutron sources operate is delineated with arrows. The error bars represent the 1-$\sigma$ uncertainties affecting the experimental Maxwellian-averaged cross sections[13, 14].

Figure 2: **KR CMa asymptotic giant branch timescale determination.**

We compare the predictions for $\log(N(Tc)/N(Zr))$ (**A**) and [Nb/Zr] (**B**) with the corresponding measured abundances (dashed lines, with statistical plus systematic errors represented as continuous lines), as a function of the time on the asymptotic giant branch (time$_{AGB}$, counted from the first thermal pulse). The maximum weight product $f_{max}$ (**C**, Methods equation 5) gives the best matching age (vertical grey line); the associated error bar (grey zone in **C**) is the age range covered by all models whose weight product is larger than $\alpha \times f_{max}$ (see Methods).

Figure 3: **Asymptotic giant branch timescales and infrared excess.**

The timescales derived from the Tc – Ru and Zr – Nb chronometers correlate well with the infrared excesses, measured by the flux ratio $R = F(12\ \mu m)/F(2.2\ \mu m)$. The dashed line marks the $R$ ratio of a star without infrared excess ($\alpha$ Tau). Error bars on the ages are computed as explained in Fig. 2; error bars on the flux ratio are computed from the statistical error on the flux density[22].

## Methods

### 1. Stellar parameters

Stellar parameters have been derived as follows. Narrow-band low-resolution spectroscopic indices, probing ZrO and TiO molecular bands, are computed according to:

$$B = 1 - \frac{\int_{\lambda_{B,i}}^{\lambda_{B,f}} F_\lambda \, d\lambda \, (\lambda_{C,f} - \lambda_{C,i})}{(\lambda_{B,f} - \lambda_{B,i}) \int_{\lambda_{C,i}}^{\lambda_{C,f}} F_\lambda \, d\lambda} \quad , \tag{1}$$

where $F_\lambda$ is the observed or model flux in the wavelength range ($\lambda$, $\lambda+d\lambda$), with the continuum window $\lambda_{C,i}$ - $\lambda_{C,f}$ and band window $\lambda_{B,i}$ - $\lambda_{B,f}$ listed in Extended Data Table 5.

These indices, along with photometric colors (Geneva $UB_1B_2V_1V_2$ and Johnson $JHK$ photometry), are calibrated against a grid of 3525 dedicated MARCS model atmospheres for S stars[10, 11] covering wide ranges in effective temperatures ($T_{eff}$ =2700 K – 4000 K), gravities (log($g$/cm s$^{-2}$) = 0 to 5), C/O ratios (C/O ≡ $N$(C)/$N$(O) = 0.500 to 0.997), s-process enrichments ([s/Fe] = 0 to 2, which reflects the zirconium enrichment), and metallicities ([Fe/H]= 0 and -0.5). Detailed synthetic spectra were calculated with state-of-the-art line lists[23] (see also Extended Data Table 5).

The comparison between synthetic and observed (photometric and band-strength) indices makes it possible[9, 11] to estimate $T_{eff}$, C/O and [s/Fe] since (i) the $(V - K)_0$ and $(J - K)_0$ colors disentangle $T_{eff}$ and C/O, and (ii) the TiO and ZrO band-strength indices disentangle $T_{eff}$ and [s/Fe]. Surface gravity is constrained by the $(U - B_1)$ and $(B_2 - V_1)$ colors, which probe bands from CN, CH, and MgH sensitive to gravity. A $\chi^2$-minimum search between the observed and synthetic indices is then performed (BMF, for "best-model finding" procedure). A set of atmospheric parameters is thus assigned to each target star (Extended Data Table 2). Because of the strong blending of atomic lines with molecular lines, it was not possible to use standard spectroscopic method to derive microturbulence, which has been fixed at $\chi_t$ = 2 km s$^{-1}$, a typical value for this kind of giants.

The validity of the BMF procedure has been assessed by two consistency checks. First, for a sample of 6 representative S stars, there is a good agreement between the C/O ratio, [Fe/H] and [s/Fe] adopted for the model atmosphere and those derived from abundance determinations performed on high-resolution spectra of the same stars[9].

Second, an accuracy test has been performed: we computed *test* model atmospheres representative of typical S stars, but with parameters ($T_{eff}$, log g, C/O, [s/Fe]) falling in between the corresponding parameter grid points. The resulting synthetic spectra were ingested in the BMF procedure as real observations. Due to the grid discretization, the model with $T_{eff}$, log g, C/O and [s/Fe] best matching the test model parameters is not always the first one selected by the code, but always lies within the first three selected models. Quantitatively, if $\chi_{best}^2$ represents the minimum $\chi^2$ difference between the test model and the first best model, the model selected by the BMF procedure is always found with a $\chi^2$ difference better than 1.4 × $\chi_{best}^2$. The error on the parameters has thus been computed as the range covered by all parameters of all models with $\chi^2$ falling in the range [$\chi_{best}^2$, 1.4 × $\chi_{best}^2$]. These parameter ranges are listed in Extended Data Table 2.

## 2. Abundance determinations

High-resolution spectra ($\lambda/\Delta\lambda$ = 85 000) were obtained with the HERMES spectrograph mounted on the 1.2m Mercator telescope at the Roque de los Muchachos Observatory (La Palma, Spain). They were reduced using the standard HERMES pipeline[8].

Abundances (Extended Data Table 3) are derived by comparing the observed spectrum and a synthetic spectrum generated by the Turbospectrum spectral-synthesis code[24], using the appropriate MARCS model atmosphere and atomic data as listed in Extended Data Table 5. Typical synthetic and observed spectra are illustrated in Extended Data Fig. 1.

The total uncertainty on these abundances can be estimated as follows:

$$\sigma_{tot}^2(A) = \sigma_A^2 + \sigma_{A,model}^2 = \sigma_A^2 + \sum_{P_i}[(\partial \log(N(A))/\partial P_i)\delta P_i]^2 \quad , \qquad (2)$$

where $N(A)$ is the number density of element A, $\sigma_A$ is the random error (line-to-line dispersion) originating from the line properties (blends, continuum location, error on oscillator strength...), $\{P_i\}$ are the model parameters, and $\delta P_i$ the typical uncertainties of the model parameters with the largest impact on the abundances: $\delta T_{eff}$ = 50 K, $\delta\log(g/\text{cm s}^{-2})$ = 0.5, $\delta\chi_t$ = 0.4 km/s. The partial derivatives $\partial \log(N(A)/\partial P_i)$ may be extracted from Extended Data Table 4, and insertion into equation 2 thus yield the following typical model errors: $\sigma_{[Zr/Fe],model}$ = 0.18, $\sigma_{[Nb/Fe],model}$ = 0.03, $\sigma_{[Nb/Zr],model}$ = 0.18, $\sigma_{Tc/Zr,model}$ = 0.29. The quadratic sum of these typical model errors and the random error specific to each star then yields the total error bar used in the abundance figures.

Extended Data Fig. 2 presents the [Zr/Fe] and [Nb/Fe] abundances measured in S and M stars. As expected, M giants display solar abundances, with some dispersion ($\pm$0.15 dex in [Zr/Fe] and [Nb/Fe]) that can be ascribed partly to uncertainties in the abundance determination (Extended Data Table 3), and partly to variations in proto-stellar cloud composition. S stars, however, segregate in Extended Data Fig. 2 along two clearly distinct sequences, which represent in fact the extrinsic/intrinsic dichotomy[12,18]. One one hand, intrinsic stars have low, close-to-solar [Nb/Fe] ratios, because $^{93}$Zr only marginally decays during the Asymptotic Giant Branch (AGB) phase. The same diagnostic applies to those carbon (C) stars (characterized by C/O> 1 in their atmosphere) flagged as AGB stars[25]. On the other hand, extrinsic stars have large Nb enhancements originating from the complete decay of $^{93}$Zr during the time elapsed since the end of the mass transfer responsible for their pollution. This segregation in terms of Nb abundance thus represents a powerful diagnostic to separate the families of intrinsic (AGB) and extrinsic (binary) stars, besides the original method based on Tc detection[12,18].

A direct confirmation of the presence of $^{93}$Zr has been attempted by searching for $^{93}$Zr lines in intrinsic S star spectra, but without success, likely because the $^{93}$Zr s-process isotopic fraction is so small ($N(^{90}\text{Zr})/N(^{91}\text{Zr})/N(^{92}\text{Zr})/N(^{93}\text{Zr})/N(^{94}\text{Zr})/N(^{96}\text{Zr})$ = 0.466 / 0.099 / 0.164 / 0.045 / 0.222 / 0.004, for a 2 M$_\odot$ solar-metallicity model after 10 thermal pulses) that the atomic isotopic shift does not induce any detectable line asymmetries. And contrary to a previous claim[26], we could not detect the $^{93}$ZrO band heads at 692.5 nm, 674.2 nm, 681.15 nm, and 639.05 nm in our high-resolution spectra.

## 3. Nucleosynthesis computations

As it ascends the AGB, a star is subject to recurrent convective instabilities that develop in the He-burning shell, the so-called thermal pulses (TP). The corresponding AGB part is dubbed 'TPAGB'. Following these He-shell flashes, the convective envelope is able to penetrate in the region previously occupied by the thermal pulse, bringing to the surface the products of a rich nucleosynthesis. This process is termed "third dredge-up" (3DUP). Nucleosynthesis by the s-process also occurs during the TPAGB, when protons from the envelope diffuse downward into the carbon-rich region of the pulse. The subsequent chain of reactions $^{12}C(p,\gamma)^{13}N(\beta^+)^{13}C(\alpha,n)^{16}O$ leads to the formation of a $^{13}C$ pocket and to the release of neutrons[1–7].

At solar metallicity, the minimum initial mass for the appearance of 3DUPs is around $M \approx 1.5 - 1.7 M_\odot$, depending on the assumed mass-loss prescription. In this paper, we computed with the STAREVOL code[27] a grid of models covering stellar masses $M_{mod} = 1.7, 2.0, 3.0, 4.0$ $M_\odot$, metallicities $Z = 0.008$ ([Fe/H]= -0.5) and $Z = 0.012$ ([Fe/H]= 0.0, i.e., solar abundances[28]), and initial [Nb/Fe] = -0.15, 0. and 0.15 dex. Both Vassiliadis & Wood (VW)[19] and Schröder (S)[20] mass-loss prescriptions were implemented. With increasing mass, the temperature at the base of the convective envelope rises and hot-bottom burning is activated. Under such conditions, corresponding to stars with $M > \sim 4$ $M_\odot$ at solar metallicity, the radiative s-process nucleosynthesis is strongly affected because the diffusing protons are burnt on the fly and the $^{13}C$ pocket is contaminated with some $^{14}N$. The result is a dramatic reduction of the production of s-elements[29]. Besides, stars with $M > \sim 4$ $M_\odot$ are able to activate the $^{22}Ne(\alpha, n)^{25}Mg$ neutron source at the base of the thermal pulse which results in an s-process signature quite different from that of lower-mass stars[30], and which, as we checked on our sample, does not agree with the measured distribution of abundances in S stars.

The s-process nucleosynthesis is obtained by post-processing[4,5,29] the stellar models with an extended nuclear network using up-to-date cross sections as available from the NETGEN nuclear library (http://www.astro.ulb.ac.be/Netgen/form.html). Large uncertainties exist on the formation of the $^{13}C$ pocket, because the efficiency of the proton mixing in the $^{12}C$ layer, responsible for the formation of that $^{13}C$ pocket, cannot be derived yet from first principles. Therefore, the extent of the partial mixing region is described by a free parameter $\lambda_{pm}$, characterizing the ratio between the extent of the proton injection region and the full intershell depth. In our computations, the values $\lambda_{pm} = 2, 3, 5, 10$ and 30% have been used. These values are larger than what would be expected from a standard overshooting prescription[4,29], but are required to match the level of observed s-process enrichments.

The measured abundances and those predicted as explained above (Extended Data Fig. 2, shaded areas) are fully consistent, with some sensitivity to the s-process modeling, mostly via the mixing parameter $\lambda_{pm}$.

## 4. Determination of TPAGB timescales

The decay of $^{93}Zr$ and $^{99}Tc$ is used to determine stellar timescales, counted from the beginning of the TPAGB (i.e., from the first thermal pulse), consistently taking into account the temperature dependence of the beta-decay rates of $^{99}Tc$ and $^{93}Zr$ in our nucleosynthesis computations[31].

It is necessary to quantify the agreement between the predicted and measured abundances of the two chronometers Tc and Nb. At time $t$ (time is counted from the first thermal pulse), the predicted Tc abundance for a star of initial mass $M_{mod}$, metallicity [Fe/H], partial mixing parameter $\lambda_{pm}$, and initial Nb abundance $N_0$(Nb), is noted $N_i$(Tc; $t$), where $i$ represents the set of model parameters $\{M_{mod}, \lambda_{pm}, N_0(Nb), [Fe/H]\}$. For each model $i$, each time $t$ is assigned a weight $W_i$ depending on the distance between $N_i$(Tc; $t$) and the observed Tc abundance. More precisely, this weight follows a Gaussian distribution which is a function of the abundance $x$ (as illustrated on the right of Extended Data Fig. 4, panels A and B):

$$W_{Tc}(x) = e^{-\frac{1}{2}\left(\frac{x-\mu(\text{Tc})}{\sigma(\text{Tc})}\right)^2}, \quad (3)$$

where $\mu(Tc)$ and $\sigma(Tc)$ are the measured Tc abundance and its associated uncertainty, respectively. A weight function $f_{Tc,i}(t)$ is computed such that, when $N_i(Tc; t^*) = x^*$, then

$$f_{\text{Tc},i}(t^*) = W_{\text{Tc}}(x^*). \quad (4)$$

A similar function $f_{Nb,i}(t)$ is computed along the same guidelines as for $f_{Tc,i}(t)$, and they are both plotted in Extended Data Fig. 4 (panel C). The weight product function, for the Tc *and* Zr chronometers, is

$$f_i(t) = f_{\text{Tc},i}(t) \times f_{\text{Nb},i}(t). \quad (5)$$

The grey curves on panel D of Extended Data Fig. 4 show this weight product $f_i(t)$ for various models. The weight product corresponding to the best fitting-model is depicted as the solid curve. Its maximum corresponds to the stellar age on the TPAGB ($f_{max} = f_{best}(t_{best})$, vertical line on panel D of Extended Data Fig. 4). The uncertainty on the TPAGB age (columns "Age$_{min}$" and "Age$_{max}$" of Table 1 and Extended Data Table 6) is computed as the time interval $[t_{min}, t_{max}]$ such that: $\forall t \in [t_{min}, t_{max}], \exists i$ such that $f_i(t) > \alpha \times f_{best}(t_{best})$, where $\alpha$ has been set to 0.6 by analogy with the 1 $\sigma$ deviation from the maximum of a Gaussian (horizontal line on panel D of Extended Data Fig. 4). The typical uncertainty on the stellar masses ($M_{mod}$ in Table 1 and Extended Data Table 6) is derived in a similar way, by considering all models having a weight product function peaking at values larger than $\alpha \times f_{best}(t_{best})$.

As an internal consistency check, we verified that the timescales derived from Tc and Nb overabundances were always consistent with the timescales derived from the minimum and maximum number of thermal pulses required to provide the measured overabundances of other s-process elements (Y, Zr, La and Ce). Furthermore, a good agreement is always found between the $^{99}$Tc and $^{93}$Zr chronometers, thus providing a well-defined TPAGB timescale (Table 1, and Extended Data Table 6), with the exception of NQ Pup, which is abnormal in other respects as well. In particular, as illustrated by its location in the HR diagram (Extended Data Fig. 3), it is the intrinsic star with the lowest luminosity[21] ($\log(L/L_\odot) = 2.96$). Its intrinsic nature is, however, secured from both its Tc and Nb content. If NQ Pup was removed from our stellar sample, it would not endanger the conclusions of the present study: TPAGB timescale correlation with IR excess (and with C/O ratio, see Table 1), would even be tighter, and the s-process operation temperature would remain unchanged.

Interestingly enough, AA Cam, with a TPAGB age of $1.4 \times 10^6$ years making it one of the least evolved TPAGB stars of our sample, has the smallest mass loss rate, as derived from CO data[32] in a sample of 10 intrinsic S stars.

Finally, we note that HR Peg is the only star of our sample displaying a strong Li line, and is presumably Li-rich. This is consistent with the fact that it is an intrinsic S star; indeed, lithium has been shown to be a useful diagnostic in the intrinsic/extrinsic classification[32]. It is believed to be produced in higher mass stars, but although present in our grid, the 4 $M_\odot$ models are never found to be compatible with the measured s-process abundances, in contrast to $M \leq 3 M_\odot$ models.

## 5. The $^{93}$Zr – $^{93}$Nb pair used as a thermometer

We first define the s-process dilution factor as:

$$f = \frac{\Delta M}{M + \Delta M} \quad , \tag{6}$$

where $M$ is the amount of mass in which $\Delta M$ is diluted. If applied to the atmosphere of an intrinsic star, $\Delta M$ is then the mass of matter freshly processed by the s-process and dredged up in the stellar envelope of mass $M$. If applied to the atmosphere of an extrinsic star, the dilution factor then combines the dilution of s-processed matter in the primary (intrinsic) star envelope, the mass-accretion efficiency and the dilution of this accreted mass in the secondary (extrinsic) star envelope.

If $X_0(\text{Zr})$, $X_s(\text{Zr})$ and $X(\text{Zr})$ are, respectively, the initial Zr mass fraction in mass $M$, the Zr mass fraction in the s-processed matter, and the final Zr mass fraction in mass $M + \Delta M$, we have:

$$X(\text{Zr}) = (1 - f) X_0(\text{Zr}) + f X_s(\text{Zr}) \quad , \tag{7}$$

and in terms of number densities ($N = X/A$, where $A$ is the mass number):

$$N(\text{Zr}) = (1 - f) N_0(\text{Zr}) + f N_s(\text{Zr}) \quad . \tag{8}$$

With the usual bracket notation:

$$\left[\frac{\text{Zr}}{\text{Fe}}\right] = \log\left(\left((1-f)\frac{N_0(\text{Zr})}{N_\odot(\text{Zr})} + f\frac{N_s(\text{Zr})}{N_\odot(\text{Zr})}\right)\frac{N_\odot(\text{Fe})}{N_s(\text{Fe})}\right) \quad . \tag{9}$$

A similar expression holds for Nb; therefore:

$$\left[\frac{\text{Zr}}{\text{Fe}}\right] = \left[\frac{\text{Nb}}{\text{Fe}}\right] + \log\left(\frac{(1-f)\frac{N_0(\text{Zr})}{N_\odot(\text{Zr})} + f\frac{N_s(\text{Zr})}{N_\odot(\text{Zr})}}{(1-f)\frac{N_0(\text{Nb})}{N_\odot(\text{Nb})} + f\frac{N_s(\text{Nb})}{N_\odot(\text{Nb})}}\right) \quad . \tag{10}$$

Asymptotically, when the abundance of the s-processed material dominates over the initial composition ($f N_s \gg (1 - f) N_0$ in equation 8):

$$\left[\frac{\text{Zr}}{\text{Fe}}\right] = \left[\frac{\text{Nb}}{\text{Fe}}\right] + \log\frac{N_s(\text{Zr})}{N_s(\text{Nb})} - \log\frac{N_\odot(\text{Zr})}{N_\odot(\text{Nb})} \quad . \tag{11}$$

We define $\omega^*$ as the ratio of the Zr to the $^{93}$Zr abundance in the s-process material:

$$\omega^* = \frac{N_s(\text{Zr})}{N_s(^{93}\text{Zr})} \quad . \tag{12}$$

When all $^{93}$Zr has decayed into mono-isotopic Nb, we have:

$$\omega^* = \frac{N_s(\text{Zr})}{N_s(\text{Nb})} \tag{13}$$

and:

$$\left[\frac{\text{Zr}}{\text{Fe}}\right] = \left[\frac{\text{Nb}}{\text{Fe}}\right] + \log\omega^* - \log\frac{N_\odot(\text{Zr})}{N_\odot(\text{Nb})} \quad . \tag{14}$$

The value $\log\omega^* - \log(N_\odot(\text{Zr})/N_\odot(\text{Nb}))$ can thus be estimated using equation 14 and Extended Data Fig. 2, by fitting a straight line of slope 1 through the points for which the approximation $fN_s \gg (1-f)N_0$ is valid. This occurs in extrinsic S stars where all $^{93}$Zr has decayed into mono-isotopic Nb and where a strong s-process pollution has erased any pristine abundance profile. Therefore, the two most Zr-rich extrinsic stars provide a measurement of $\omega^*$ because for these stars $N(\text{Zr})/N(\text{Nb}) \approx N_s(\text{Zr})/N_s(\text{Nb}) \equiv \omega^*$.

On the other hand, nuclear physics provides the temperature dependence of $\omega^*$: it is well known that, to a very good approximation, the s-process flow is in local equilibrium within a given isotopic chain. In the case of Zr, this is true at least between $^{90}$Zr and $^{94}$Zr where no branching point occurs, so that $N_s(^A\text{Zr}) \times \langle\sigma_A\rangle$ is constant within the Zr chain ($\langle\sigma_A\rangle$ is the Maxwellian-averaged neutron-capture cross section[13, 14] of $^A$Zr). Thus:

$$\omega^* = \langle\sigma_{93}\rangle \times \left[\frac{1}{\langle\sigma_{90}\rangle} + \frac{1}{\langle\sigma_{91}\rangle} + \frac{1}{\langle\sigma_{92}\rangle} + \frac{1}{\langle\sigma_{94}\rangle}\right] \quad . \tag{15}$$

Since $\omega^*$ is temperature-dependent through the temperature dependence of the individual cross sections, it can be used to constrain the temperature at which the Zr production took place.

We find a temperature for the s-process less than $\sim 2.5 \times 10^8$ K, favoring the $^{13}$C($\alpha$,n)$^{16}$O neutron source with respect to the $^{22}$Ne($\alpha$,n)$^{25}$Mg one. The deduced temperature range (Fig. 1) is an upper estimate: if we had considered more stars as being dominated by the s-process pollution, the y-intercept of equation 14 (i.e., $\log\omega^* - \log(N_\odot(\text{Zr})/N_\odot(\text{Nb}))$) would have been larger, and the derived temperature even lower (see Fig. 1), strengthening the case for the $^{13}$C($\alpha$,n)$^{16}$O neutron source. Finally, the derived temperature is independent of the solar abundances and their associated errors, as can be seen from equation 14, because the bracket notation corresponds to abundance ratios normalized to their solar values, and these cancel out.

Taking mass loss into account in the dilution factor (Eq. 6) will increase $f$, so the condition $fN_s \gg (1-f)N_0$ will be reached even sooner. Since only this asymptotic regime is used to constrain the s-process operation temperature, our temperature determination is unaffected by a diminution of the envelope mass.

Previous estimates of the s-process temperature were based on the analysis of solar system abundances[1,28], essentially derived from primitive carbonaceous chondrite meteorites of type C1, where most s-process branchings lead to a high temperature of the order of $3 \times 10^8$ K for the s-process operation, thus pointing toward the activation of the $^{22}$Ne($\alpha$,n)$^{25}$Mg neutron source. The higher temperatures so inferred were ascribed to a short, final neutron burst from the $^{22}$Ne($\alpha$,n)$^{25}$Mg neutron source operating in a late, hot thermal pulse. However, these meteoritic abundances are notoriously difficult to interpret[1, 2, 33] since: (i) they are a mix of diverse nucleosynthetic events taking place in stars of different masses and metallicities, (ii) they are potentially sensitive to differential depletion effects during grain condensation and contamination with solar-system material[34]; (iii) the temperatures are deduced from branchings (i.e., unstable nuclei for which there is competition between neutron-capture and $\beta$-decay) along the s-process path. However some branchings are not only sensitive to temperature, but also to poorly-known neutron and electron densities; (iv) the branching neutron-capture cross-sections and the temperature dependence of the $\beta$-decays are not always well constrained experimentally.

Our work shows that the event associated with $^{22}$Ne($\alpha$,n)$^{25}$Mg, despite imprinting its signature in the material incorporated in the solar-system carbonaceous chondrites, is not representative of the s-process in low-mass stars, which is powered by the $^{13}$C($\alpha$,n)$^{16}$O neutron source.

# Extended data table legends

Extended Data Table 1: **Observations summary.**

The wavelength range of the HERMES spectrograph[8] is 380 – 900 nm. T is the exposure time; the signal-to-noise ratio (S/N) is estimated around 500 nm.

Extended Data Table 2: **Adopted atmospheric parameters for the S stars of the present study.**

The "min" and "max" columns list the parameter range covered by the various best-fitting model atmospheres (see text). The last column lists $\chi_{best}^2$, i.e. the minimum $\chi^2$ difference between spectral and photometric indices computed on the observed spectrum and on a synthetic spectrum corresponding to the listed parameters.

Extended Data Table 3: **Derived abundances for S, M and $C^{25}$ stars.**

The S-star spectral types are from the General Catalog of S Stars, second edition[35]. The usual spectroscopic notation is adopted: for elements A and B, $\log \epsilon_A = \log(N(A)/N(H)) + 12.0$ and $[A/B]= \log(N(A)/N(B)) - \log(N(A)/N(B))_\odot$. The e/i column indicates whether the star abundance peculiarities are of extrinsic (e) or intrinsic (i) origin.

Extended Data Table 4: **Sensitivity of Fe, Nb, Zr and Tc abundances on model atmosphere parameter variations (effective temperature, gravity and microturbulence $\chi_t$).**

The nominal model atmosphere has $T_{eff}$ = 3500 K, $\log(g/\text{cm s}^{-2})$ = 1, $\chi_t$ = 2 km s$^{-1}$, [Fe/H] = -0.5 dex, C/O = 0.752 and [s/Fe] = 1.00 dex.

Extended Data Table 5: **Molecular and atomic lines used for computation of the stellar parameters, and atomic lines used for the abundance determinations.**

**a**, Boundaries (expressed in nm) of the continuum ($\lambda_{C,i}, \lambda_{C,f}$) and band ($\lambda_{B,i}, \lambda_{B,f}$) windows used in the computation of the band indices (Equation 1). **b**, Atomic line data (wavelength, excitation potential and oscillator strength[23]) for the elements with abundances derived in the present study. For Zr, lines were restricted to those two lines having laboratory oscillator strengths[36]. The hyperfine structure of Tc is included.

Extended Data Table 6: **Derived s-process timescales, counted from the first thermal pulse, and masses for the intrinsic S stars of our sample.**

The adopted mass-loss prescription is indicated by 'VW' (Vassiliadis & Wood) or 'S' (Schröder). The TP column indicates the number of thermal pulses that took place in the star, up to the present time. The WP column lists the maximum of the weight product function

(computed from equation 5). $M_{mod}$ and $M_{HR}$ stand for the mass derived from nucleosynthesis models and stellar-evolution models (i.e., location in the Hertzsprung-Russell diagram, Extended Data Fig. 3), respectively. The partial mixing parameter $\lambda_{pm}$ expresses the mixing depth of protons into the C-rich intershell zone, in units of the width of the intershell zone, as required to activate the $^{13}C(\alpha,n)^{16}O$ neutron source. The column labelled 'C/O' provides an estimate of the C/O ratio (selected among the grid points C/O = 0.5, 0.75, 0.9, 0.92, 0.97, 0.99) best matching the spectroscopic and photometric data. The infrared excess $R = F(12\ \mu m)/F(2.2\ \mu m)$ is listed in the last column.

**Extended data figure legends**

Extended Data Figure 1: **Observed and synthetic spectra around the Tc I 426.227 nm line.**

Panel A shows the observed spectrum of the intrinsic, Tc-rich star NQ Pup (dots) along with spectral synthesis with $\log(\epsilon_{Tc}) = -0.20$ (continuous line, see Extended Data Table 3) and no Tc (dashed line). Panel B: same as panel A for the extrinsic, Tc-poor star V613 Mon; spectral synthesis correspond to $\log(\epsilon_{Tc}) = -0.30$ (continuous line) and no Tc (dashed line).

Extended Data Figure 2: **Predicted versus measured abundances.**

We show that Nb and Zr abundances for intrinsic S stars and carbon stars (with only upper limits on the Nb abundance[25]) are consistent with nucleosynthesis predictions (shaded areas) for a 2.0 $M_\odot$ model with a solar initial Nb abundance ($\pm 0.15$ dex) and different $\lambda_{pm}$[29] (as labelled). Once all $^{93}$Zr has decayed to $^{93}$Nb, the models (dotted lines) reproduce the extrinsic S stars. The error bars represent the line-to-line dispersion and the systematic uncertainty from the models. The dashed line is a fit of equation 14 through the two most enriched extrinsic stars.

Extended Data Figure 3: **Location of intrinsic S stars in the Hertzsprung-Russell diagram.**

We show the location of intrinsic S stars (triangles) from our sample in the stellar luminosity versus effective temperature diagram[21], along with low-mass evolutionary tracks (starting at the zero-age main sequence). The main sequence (MS) is indicated. The black diamond on each evolutionary track indicates where the "third dredge-up" events first start.

Extended Data Figure 4: **Timescale determination for HIP 103476.**

For clarity, only $N(Tc)/N(Zr)$ and [Nb/Zr] predictions from the best-matching model (as judged from panel **D**), are plotted in panels **A** and **B** (same meaning as in Fig. 2). The deviation between predicted (solid line) and observed (dashed line) abundances is converted into weight functions $f_{Tc,i}(t)$ and $f_{Nb,i}(t)$ (equations 4, panel **C**). The maximum value reached by their product (equation 5, solid curve in **D**) provides the best age estimate (vertical line in **D**). Weight-product functions corresponding to less-satisfactory models are plotted as grey lines in panel **D**.

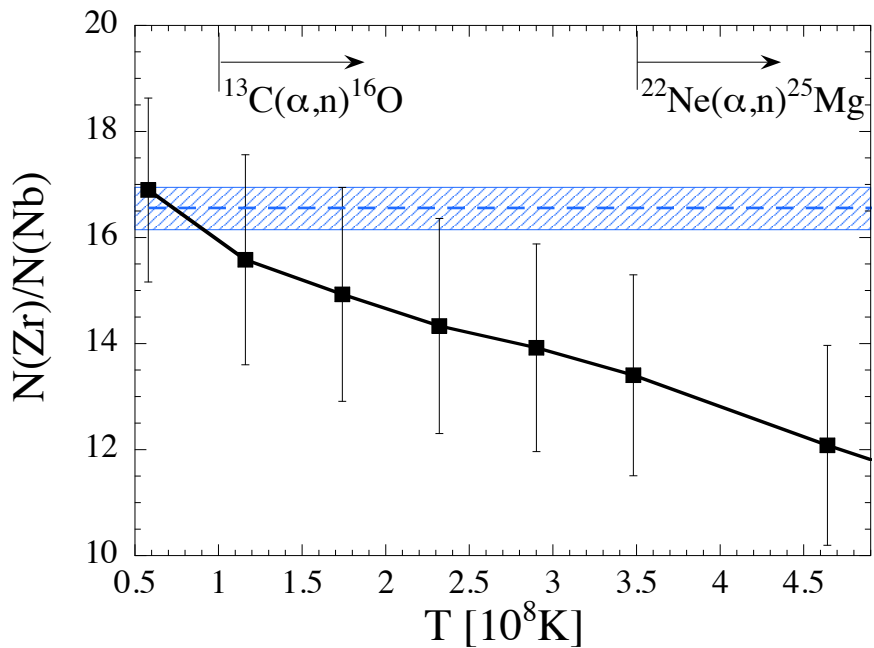

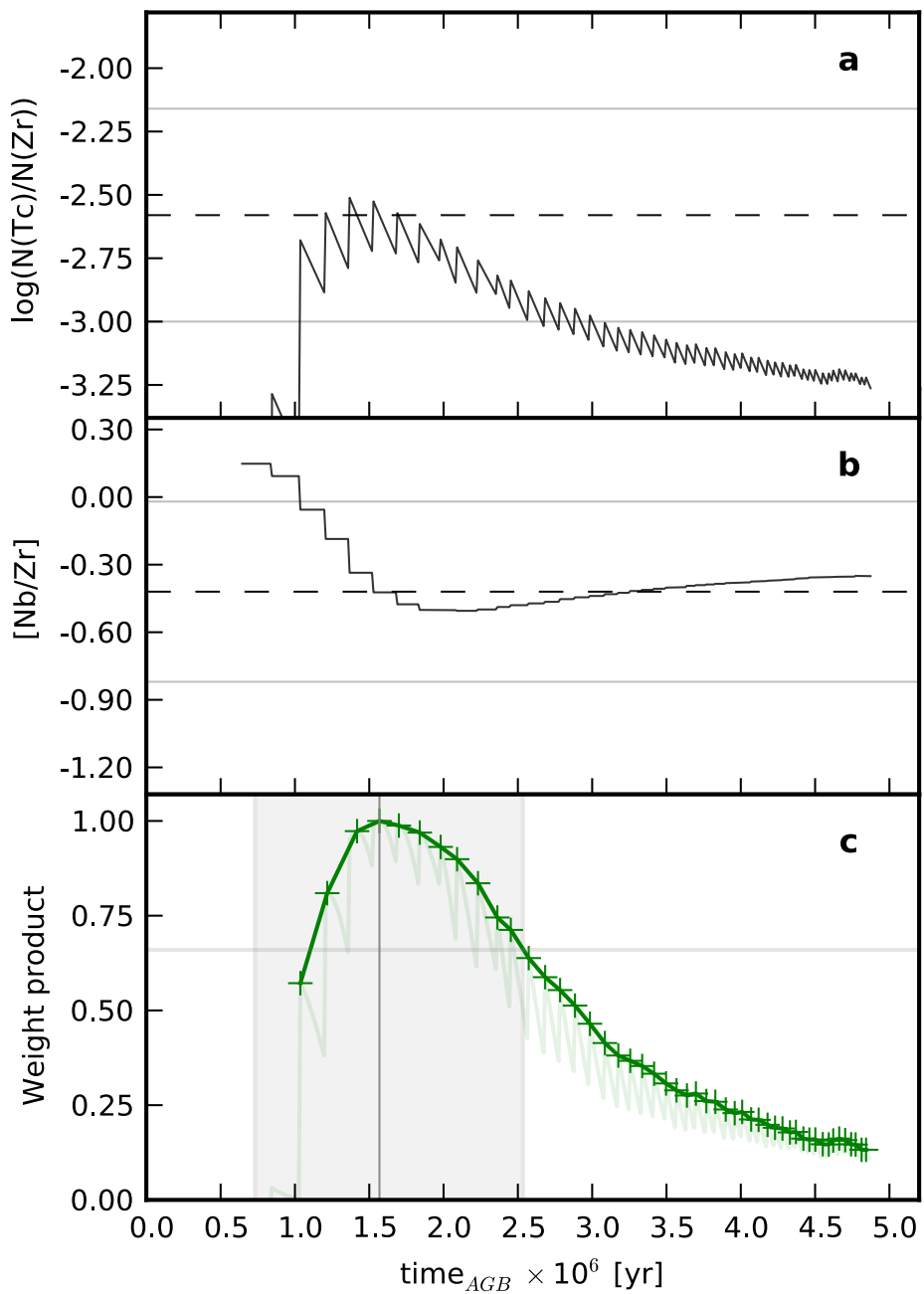

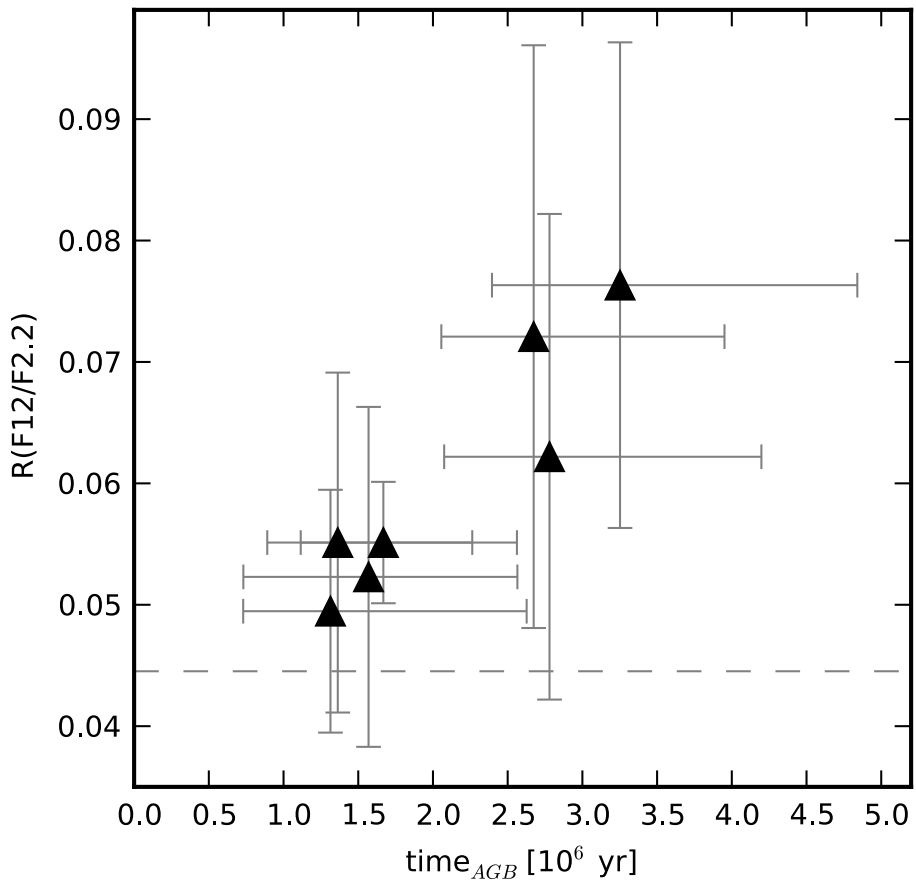

| Star name | Date Obs. | T (s) | S/N |
|---|---|---|---|
| CSS 454 | 2011, Feb 17 | 2800 | 40 |
| $o^1$ Ori | 2011, Apr 10 | 60 | 74 |
| KR CMa | 2011, Apr 6 | 1200 | 50 |
| NQ Pup | 2010, Feb 20 | 800 | 100 |
| AD Cyg | 2009, May 19 | 2000 | 78 |
| HR Peg | 2009, Jul 18 | 420 | 44 |
| AA Cam | 2010, Apr 3 | 600 | 39 |
| HIP 103476 | 2009, Jun 20 | 1500 | 130 |
| BD Cam | 2009, Sep 17 | 120 | 109 |
| HD 119667 | 2011, Apr 10 | 995 | 57 |
| V613 Mon | 2009, Sep 24 | 800 | 50 |
| HD 191226 | 2009, May 19 | 500 | 78 |
| HD 191589 | 2011, Apr 7 | 533 | 55 |
| V530 Lyr | 2011, Apr 2 | 647 | 39 |
| CPD -19°1672 | 2011, Jan 11 | 2600 | 50 |
| HR 363 | 2009, Jul 27 | 200 | 84 |
| V1261 Ori | 2009, Nov 29 | 500 | 55 |
| RR UMi | 2010, Feb 7 | 300 | 78 |
| V 762 Cas | 2011, Aug 11 | 400 | 110 |
| V 465 Cas | 2010, Aug 23 | 400 | 100 |
| $\mu$ Gem | 2012, Mar 9 | 900 | 212 |
| $\rho$ Per | 2012, Jul 8 | 60 | 141 |
| RZ Ari | 2013, Feb 9 | 100 | 39 |

| Star | $T_{\text{eff}}$ (K) | | | $\log(g/\text{cm s}^{-2})$ | | | [Fe/H] | | | C/O | | | [s/Fe] | | | $\chi^2_{\text{best}}$ |
|---|---|---|---|---|---|---|---|---|---|---|---|---|---|---|---|---|
| | | min | max | | min | max | | min | max | | min | max | | min | max | |
| CSS 454 | 3500 | 3500 | 3500 | 0. | 0. | 0. | -0.5 | -0.5 | -0.5 | 0.50 | 0.50 | 0.50 | 1. | 1. | 1. | 0.0278 |
| $o^1$ Ori | 3600 | 3500 | 3600 | 0. | 0. | 1. | 0.0 | -0.5 | 0.0 | 0.50 | 0.50 | 0.92 | 0. | 0. | 1. | 0.0911 |
| KR CMa | 3400 | 3300 | 3400 | 0. | 0. | 1. | 0.0 | -0.5 | 0.0 | 0.50 | 0.50 | 0.75 | 0. | 0. | 0. | 0.0392 |
| NQ Pup | 3500 | 3500 | 3500 | 1. | 0. | 1. | -0.5 | -0.5 | -0.5 | 0.50 | 0.50 | 0.75 | 1. | 1. | 1. | 0.0193 |
| AD Cyg | 3300 | 3100 | 3300 | 0. | 0. | 1. | 0.0 | -0.5 | 0.0 | 0.97 | 0.97 | 0.99 | 1. | 1. | 1. | 0.1450 |
| HR Peg | 3500 | 3500 | 3500 | 1. | 1. | 2. | -0.5 | -0.5 | -0.5 | 0.75 | 0.50 | 0.75 | 1. | 1. | 1. | 0.0187 |
| AA Cam | 3400 | 3300 | 3400 | 1. | 1. | 2. | 0.0 | -0.5 | 0.0 | 0.50 | 0.50 | 0.75 | 0. | 0. | 0. | 0.0233 |
| HIP 103476 | 3400 | 3400 | 3500 | 1. | 0. | 1. | -0.5 | -0.5 | -0.5 | 0.50 | 0.50 | 0.75 | 1. | 1. | 1. | 0.0628 |
| BD Cam | 3600 | 3600 | 3600 | 1. | 0. | 1. | 0.0 | 0.0 | 0.0 | 0.50 | 0.50 | 0.50 | 1. | 1. | 1. | 0.0216 |
| HD 119667 | 3700 | 3700 | 3700 | 1. | 1. | 2. | -0.5 | -0.5 | -0.5 | 0.50 | 0.50 | 0.50 | 1. | 1. | 1. | 0.0200 |
| V613 Mon | 3500 | 3500 | 3500 | 0. | 0. | 1. | -0.5 | -0.5 | -0.5 | 0.75 | 0.75 | 0.95 | 1. | 1. | 1. | 0.0987 |
| HD 191226 | 3600 | 3500 | 3700 | 1. | 0. | 1. | -0.5 | -0.5 | 0.0 | 0.75 | 0.50 | 0.99 | 0. | 0. | 1. | 0.2941 |
| HD 191589 | 3700 | 3600 | 3800 | 1. | 0. | 1. | 0.0 | -0.5 | 0.0 | 0.50 | 0.50 | 0.97 | 0. | 0. | 0. | 0.0462 |
| V530 Lyr | 3500 | 3500 | 3600 | 0. | 0. | 0. | -0.5 | -0.5 | 0.0 | 0.50 | 0.50 | 0.75 | 0. | 0. | 0. | 0.0510 |
| CPD -19°1672 | 3600 | 3600 | 3700 | 1. | 0. | 1. | -0.5 | -0.5 | 0.0 | 0.75 | 0.50 | 0.95 | 0. | 0. | 0. | 0.0404 |
| HR 363 | 3600 | 3600 | 3600 | 1. | 0. | 1. | -0.5 | -0.5 | -0.5 | 0.50 | 0.50 | 0.50 | 1. | 1. | 1. | 0.0351 |
| V1261 Ori | 3600 | 3600 | 3700 | 1. | 1. | 2. | -0.5 | -0.5 | 0.0 | 0.50 | 0.50 | 0.75 | 1. | 1. | 1. | 0.0558 |
| RR UMi | 3500 | 3400 | 3500 | 1. | 1. | 3. | 0.0 | 0.0 | 0.0 | 0.50 | 0.50 | 0.50 | 0. | 0. | 0. | 0.0154 |
| V762 Cas | 3800 | 3700 | 3800 | 0. | 0. | 0. | 0.0 | -0.5 | 0.0 | 0.50 | 0.50 | 0.95 | 0. | 0. | 0. | 0.0199 |
| V465 Cas | 3400 | 3300 | 3400 | 1. | 2. | 2. | -0.5 | -0.5 | -0.5 | 0.50 | 0.50 | 0.50 | 0. | 0. | 0. | 0.0146 |
| $\mu$ Gem | 3600 | 3600 | 3600 | 1. | 0. | 1. | 0.0 | 0.0 | 0.0 | 0.50 | 0.50 | 0.75 | 0. | 0. | 0. | 0.0807 |
| $\rho$ Per | 3500 | 3500 | 3500 | 1. | 0. | 1. | -0.5 | -0.5 | -0.5 | 0.50 | 0.50 | 0.50 | 0. | 0. | 0. | 0.0365 |
| RZ Ari | 3400 | 3300 | 3500 | 0. | 0. | 2. | -0.5 | -0.5 | 0.0 | 0.50 | 0.50 | 0.75 | 0. | 0. | 0. | 0.1212 |

| Star | Spec. Typ. | e/i | [Fe/H] | $\sigma_{\rm[Fe/H]}$ | [Nb/Fe] | $\sigma_{\rm[Nb/Fe]}$ | [Zr/Fe] | $\sigma_{\rm[Zr/Fe]}$ | $\log \epsilon_{\rm Tc}$ | $\sigma(\log \epsilon_{\rm Tc})$ |
|---|---|---|---|---|---|---|---|---|---|---|
| CSS 454 | S | i | -0.40 | 0.12 | -0.07 | 0.24 | 0.80 | 0.21 | 0.20 | 0.20 |
| $o^1$ Ori | M3S | i | -0.45 | 0.07 | -0.04 | 0.17 | 0.45 | 0.20 | -0.05 | 0.20 |
| KR CMa | M4S | i | -0.34 | 0.11 | 0.02 | 0.23 | 0.44 | 0.27 | 0.10 | 0.20 |
| NQ Pup | S5/2 | i | -0.31 | 0.13 | -0.11 | 0.16 | 0.76 | 0.21 | -0.20 | 0.20 |
| AD Cyg | S5/5 | i | -0.05 | 0.14 | 0.05 | 0.24 | 0.90 | 0.20 | 0.30 | 0.20 |
| HR Peg | S4/1 | i | 0.00 | 0.06 | 0.06 | 0.08 | 0.55 | 0.18 | -0.10 | 0.10 |
| AA Cam | M5S | i | -0.04 | 0.05 | 0.02 | 0.24 | 0.15 | 0.25 | 0.10 | 0.20 |
| HIP 103476 | MS | i | -0.01 | 0.06 | -0.03 | 0.16 | 0.61 | 0.18 | 0.30 | 0.20 |
| BD Cam | S4/2 | e | -0.03 | 0.06 | 0.58 | 0.16 | 0.78 | 0.18 | - | - |
| HD 119667 | M1wkS | e | 0.01 | 0.07 | 0.46 | 0.17 | 0.59 | 0.18 | - | - |
| V613 Mon | S3/2 | e | -0.26 | 0.15 | 0.42 | 0.22 | 0.66 | 0.21 | - | - |
| HD 191226 | M1-3S | e | -0.28 | 0.10 | 0.31 | 0.19 | 0.35 | 0.16 | - | - |
| HD 191589 | S | e | 0.01 | 0.06 | 0.29 | 0.17 | 0.29 | 0.15 | - | - |
| V530 Lyr | S3/1 | e | -0.16 | 0.09 | 0.19 | 0.23 | 0.29 | 0.20 | - | - |
| CPD -19°1672 | S3 | e | -0.01 | 0.06 | 0.37 | 0.19 | 0.41 | 0.21 | - | - |
| HR 363 | S3/2 | e | -0.38 | 0.12 | 0.77 | 0.19 | 0.83 | 0.22 | - | - |
| V1261 Ori | S4,1 | e | -0.22 | 0.15 | 0.81 | 0.23 | 0.85 | 0.22 | - | - |
| RR UMi | M5III | | -0.06 | 0.06 | -0.01 | 0.13 | -0.04 | 0.22 | - | - |
| V762 Cas | M3 | | -0.06 | 0.09 | -0.04 | 0.17 | -0.01 | 0.14 | - | - |
| V465 Cas | M3 | | 0.03 | 0.05 | 0.03 | 0.27 | -0.13 | 0.21 | - | - |
| $\mu$ Gem | M3III | | -0.15 | 0.06 | 0.18 | 0.18 | 0.05 | 0.20 | - | - |
| $\rho$ Per | M4II | | -0.06 | 0.09 | 0.10 | 0.17 | -0.09 | 0.23 | - | - |
| RZ Ari | M6III | | -0.22 | 0.20 | -0.15 | 0.27 | 0.02 | 0.25 | - | - |
| [IGI95] C1 | C | | -0.8 | 0.3 | ≤0.10 | – | 0.50 | 0.30 | - | - |
| [IGI95] C3 | C | | -0.5 | 0.3 | ≤0.30 | – | 1.20 | 0.30 | - | - |
| BMB-B 30 | C | | -1.0 | 0.3 | ≤0.00 | – | 0.40 | 0.30 | - | - |

| Element | $\Delta\,T_{\text{eff}} = +100$ K | $\Delta \log(g/\text{cm s}^{-2}) = +0.5$ | $\Delta\chi_t = -0.4$ (km s$^{-1}$) |
|---|---|---|---|
| $\Delta \log N(\text{Fe})$ | -0.10 | +0.14 | +0.07 |
| $\Delta \log N(\text{Nb})$ | -0.15 | +0.15 | +0.07 |
| $\Delta \log N(\text{Zr})$ | -0.13 | +0.04 | -0.08 |
| $\Delta \log N(\text{Tc})$ | +0.10 | +0.30 | +0.00 |
| $\Delta$ [Zr/Fe] | -0.03 | -0.10 | -0.15 |
| $\Delta$ [Nb/Fe] | -0.05 | +0.01 | +0.00 |
| $\Delta$ [Nb/Zr] | -0.02 | +0.11 | +0.15 |
| $\Delta \log(N(\text{Tc})/N(\text{Zr}))$ | +0.23 | +0.26 | +0.08 |

**a**

| Band | $\lambda_{B,i}$ | $\lambda_{B,f}$ | $\lambda_{C,i}$ | $\lambda_{C,f}$ |
|---|---|---|---|---|
| ZrO | 574.8 | 575.7 | 580.0 | 581.0 |
| ZrO | 634.2 | 637.3 | 646.2 | 647.2 |
| ZrO | 637.8 | 638.2 | 646.2 | 647.2 |
| ZrO | 650.5 | 653.0 | 646.2 | 647.2 |
| TiO | 544.8 | 545.4 | 541.6 | 544.7 |
| TiO | 559.1 | 560.0 | 541.6 | 544.7 |
| TiO | 561.5 | 562.0 | 580.0 | 581.0 |
| TiO | 668.1 | 670.6 | 646.2 | 647.2 |
| TiO | 671.4 | 673.5 | 646.2 | 647.2 |
| TiO | 705.4 | 706.9 | 703.0 | 705.0 |
| TiO | 712.5 | 714.4 | 703.0 | 705.0 |
| Na | 588.3 | 590.3 | 580.0 | 581.0 |

**b**

| Element | $\lambda$ (nm) | $\chi_{exc}(eV)$ | $\log(gf)$ | Element | $\lambda$ (nm) | $\chi_{exc}(eV)$ | $\log(gf)$ |
|---|---|---|---|---|---|---|---|
| Fe I | 872.914 | 3.41 | -2.87 | Tc I | 426.2297 | 0.00 | -1.297 |
| | 871.039 | 4.91 | -0.54 | | 426.2320 | 0.00 | -1.297 |
| | 869.870 | 2.99 | -3.45 | | 426.2324 | 0.00 | -2.654 |
| | 833.191 | 4.38 | -0.54 | | 426.2327 | 0.00 | -1.353 |
| | 810.832 | 2.72 | -3.89 | | 426.2346 | 0.00 | -1.353 |
| | 792.414 | 4.76 | -1.68 | | 426.2348 | 0.00 | -2.541 |
| | 748.174 | 2.75 | -4.10 | | 426.2350 | 0.00 | -1.549 |
| | 746.152 | 2.55 | -3.58 | | 426.2364 | 0.00 | -1.549 |
| | 744.302 | 4.18 | -1.82 | | 426.2366 | 0.00 | -1.590 |
| | 743.053 | 2.58 | -3.86 | Ce II | 876.991 | 0.55 | -2.37 |
| | 735.350 | 4.73 | -1.57 | | 875.799 | 0.59 | -2.29 |
| Y I | 880.059 | 0.00 | -2.24 | | 872.131 | 0.33 | -2.48 |
| | 734.646 | 2.01 | -0.82 | | 870.237 | 0.47 | -1.46 |
| Y II | 788.188 | 1.84 | -0.57 | | 870.075 | 0.46 | -2.77 |
| | 745.028 | 1.75 | -1.59 | | 840.525 | 0.30 | -2.10 |
| Zr I | 784.937 | 0.69 | -1.30 | | 840.413 | 0.70 | -1.67 |
| | 781.937 | 1.82 | -0.38 | | 839.451 | 0.27 | -2.59 |
| Nb I | 411.689 | 0.00 | -1.18 | | 826.405 | 0.00 | -2.56 |
| | 419.509 | 0.02 | -0.91 | | 826.342 | 0.33 | -2.24 |
| | 434.531 | 0.00 | -1.36 | | 824.047 | 0.32 | -2.21 |
| | 424.946 | 0.00 | -1.47 | | 790.926 | 0.32 | -2.43 |
| | 426.205 | 0.13 | -0.56 | | 789.897 | 0.90 | -1.12 |
| | 535.072 | 0.27 | -0.86 | | 751.876 | 0.46 | -2.94 |
| Tc I | 426.2215 | 0.00 | -0.879 | La II | 480.900 | 0.24 | -1.40 |
| | 426.2219 | 0.00 | -1.527 | | 480.404 | 0.24 | -1.49 |
| | 426.2252 | 0.00 | -1.527 | | 474.873 | 0.93 | -0.54 |
| | 426.2257 | 0.00 | -1.191 | | 466.249 | 0.00 | -2.25 |
| | 426.2261 | 0.00 | -1.337 | | 455.846 | 0.32 | -0.97 |
| | 426.2290 | 0.00 | -1.337 | | 433.375 | 0.17 | -0.06 |
| | 426.2293 | 0.00 | -1.655 | | 432.246 | 0.17 | -2.38 |

| Star name | Mass loss | Age (Myr) | TP | Age$_{min}$ (Myr) | Age$_{max}$ (Myr) | WP | $M_{mod}$ ($M_\odot$) | $\lambda_{pm}$ | $\lambda_{pm}^{min}$ | $\lambda_{pm}^{max}$ | $M_{HR}$ ($M_\odot$) | $\log L$ ($L_\odot$) | C/O | $R$ |
|---|---|---|---|---|---|---|---|---|---|---|---|---|---|---|
| $o^1$ Ori | VW | 1.3 | 9 | 0.7 | 2.6 | 1.00 | 2 | 10% | 2% | 30% | 2.5 | 3.45 | 0.50 | 0.049 |
|  | S | 1.8 | 9 | 0.7 | 3.9 | 1.00 | 3 | 5% | 2% | 30% |  |  |  |  |
| AA Cam | VW | 1.4 | 8 | 0.9 | 2.3 | 0.99 | 3 | 5% | 2% | 10% | 3.0 | 3.91 | 0.50 | 0.055 |
|  | S | 1.6 | 8 | 0.7 | 2.7 | 1.00 | 3 | 5% | 2% | 10% |  |  |  |  |
| KR CMa | VW | 1.6 | 10 | 0.7 | 2.6 | 1.00 | 3 | 10% | 2% | 30% | – | – | 0.50 | 0.052 |
|  | S | 1.8 | 9 | 0.7 | 3.9 | 1.00 | 3 | 10% | 2% | 30% |  |  |  |  |
| CSS 454 | VW | 1.7 | 12 | 1.1 | 2.6 | 0.88 | 2 | 10% | 5% | 10% | – | – | 0.50 | 0.055 |
|  | S | 1.7 | 12 | 1.0 | 3.9 | 0.88 | 2 | 10% | 5% | 30% |  |  |  |  |
| HIP 103476 | VW | 2.2 | 14 | 1.2 | 3.2 | 1.00 | 3 | 10% | 5% | 10% | 2.0 | 3.59 | 0.50 | – |
|  | S | 2.5 | 13 | 1.5 | 4.0 | 1.00 | 3 | 10% | 5% | 10% |  |  |  |  |
| AD Cyg | VW | 2.7 | 18 | 2.1 | 3.9 | 0.57 | 3 | 10% | 3% | 10% | – | – | 0.97 | 0.072 |
|  | S | 3.0 | 17 | 2.2 | 4.0 | 0.53 | 3 | 10% | 3% | 10% |  |  |  |  |
| NQ Pup | VW | 2.8 | 19 | 2.1 | 4.2 | 0.28 | 3 | 3% | 2% | 3% | 1.5 | 2.95 | 0.50 | 0.062 |
|  | S | 3.1 | 18 | 2.3 | 4.0 | 0.33 | 3 | 10% | 3% | 10% |  |  |  |  |
| HR Peg | VW | 3.2 | 23 | 2.4 | 4.8 | 0.94 | 3 | 3% | 2% | 5% | 2.0 | 3.39 | 0.75 | 0.076 |
|  | S | 3.0 | 17 | 2.4 | 3.7 | 0.73 | 3 | 3% | 3% | 10% |  |  |  |  |

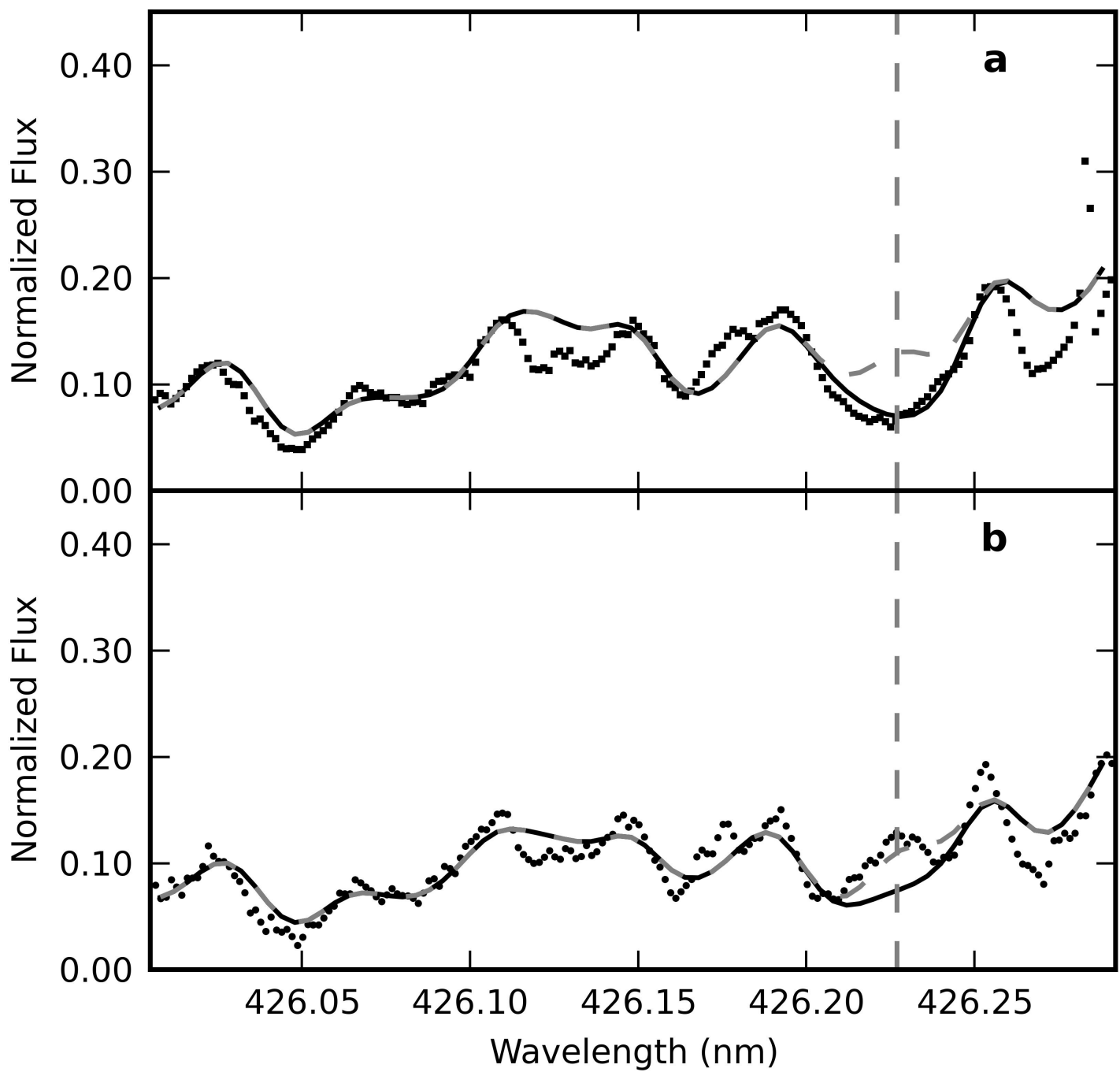

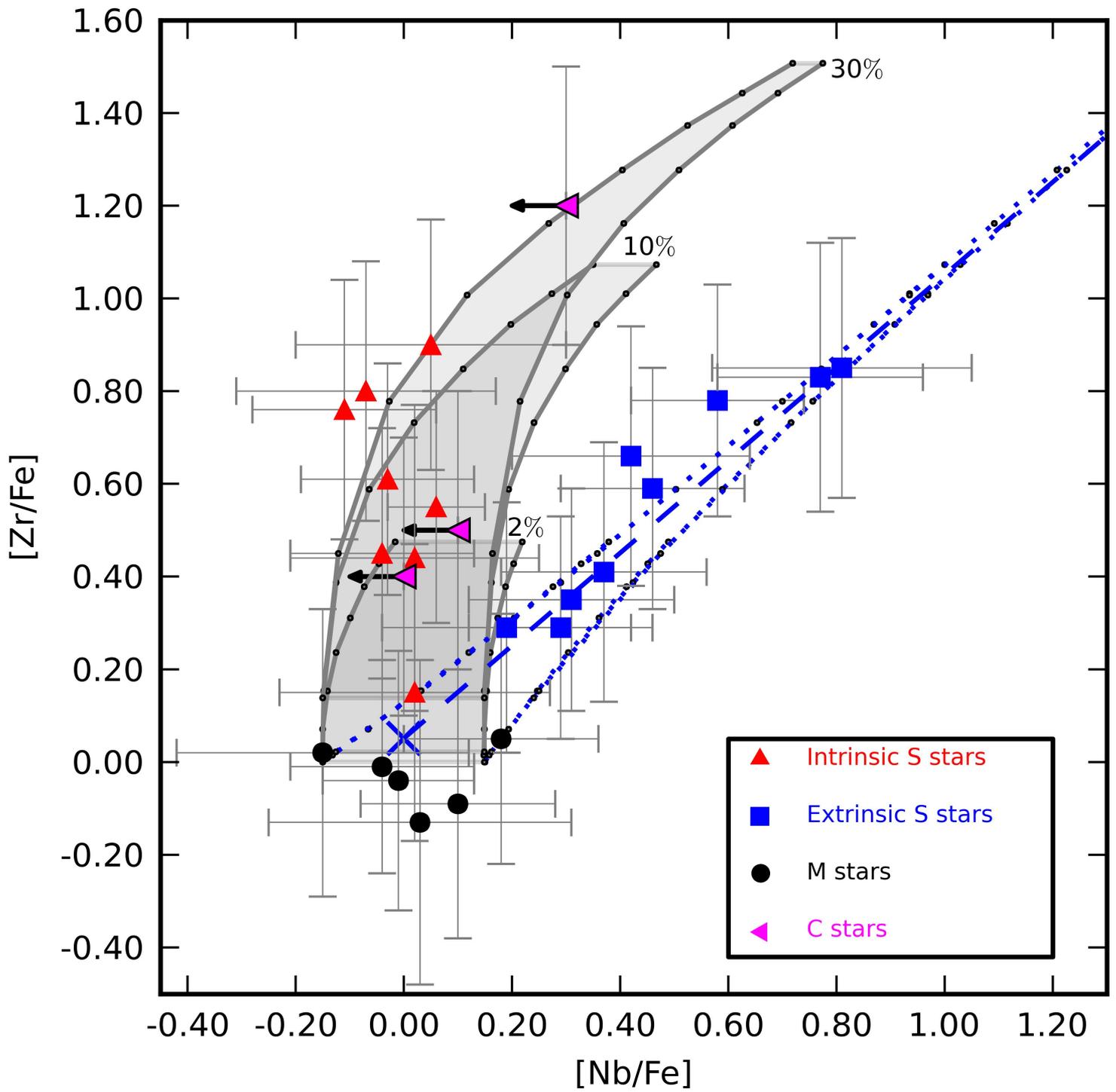

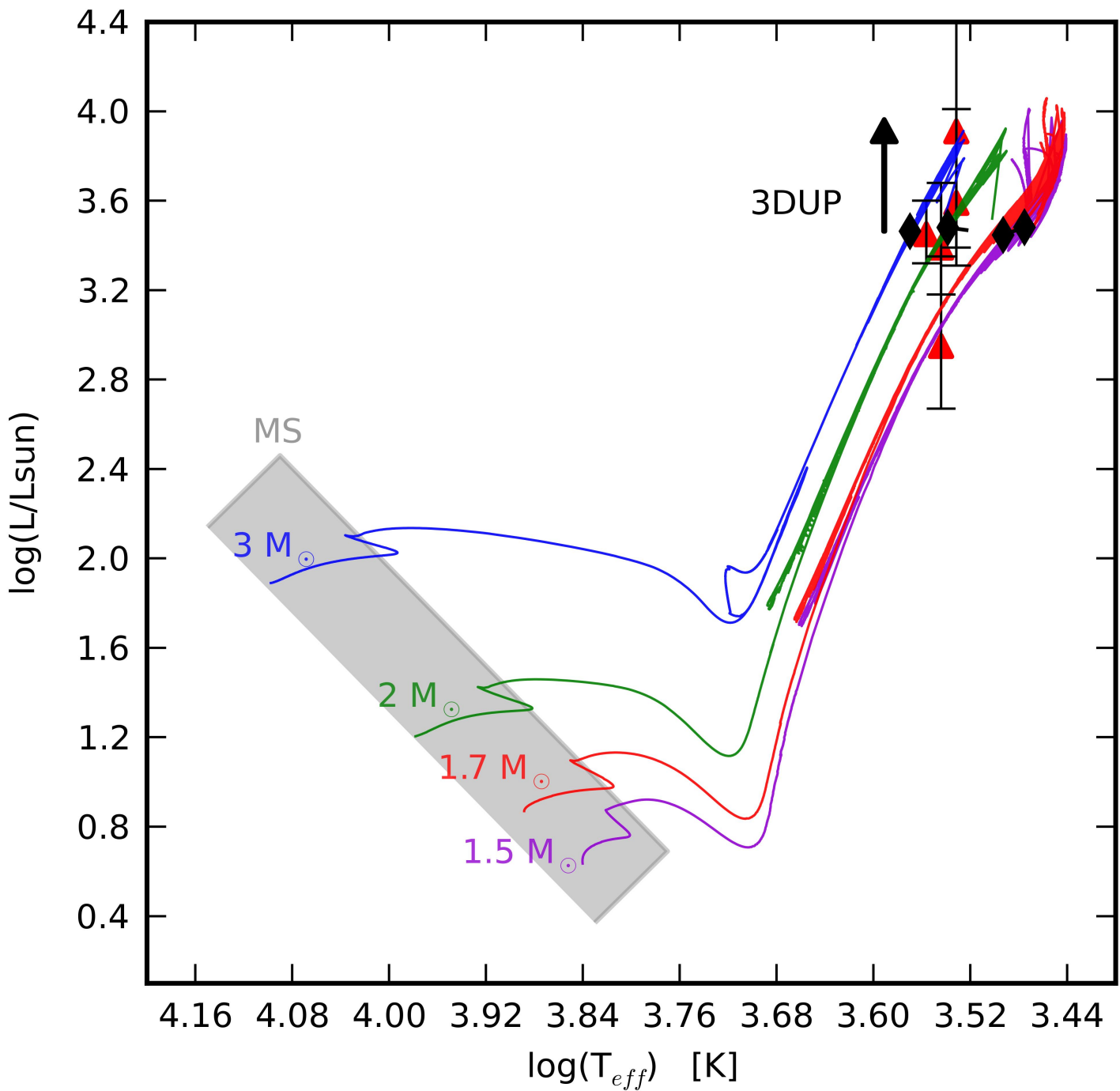

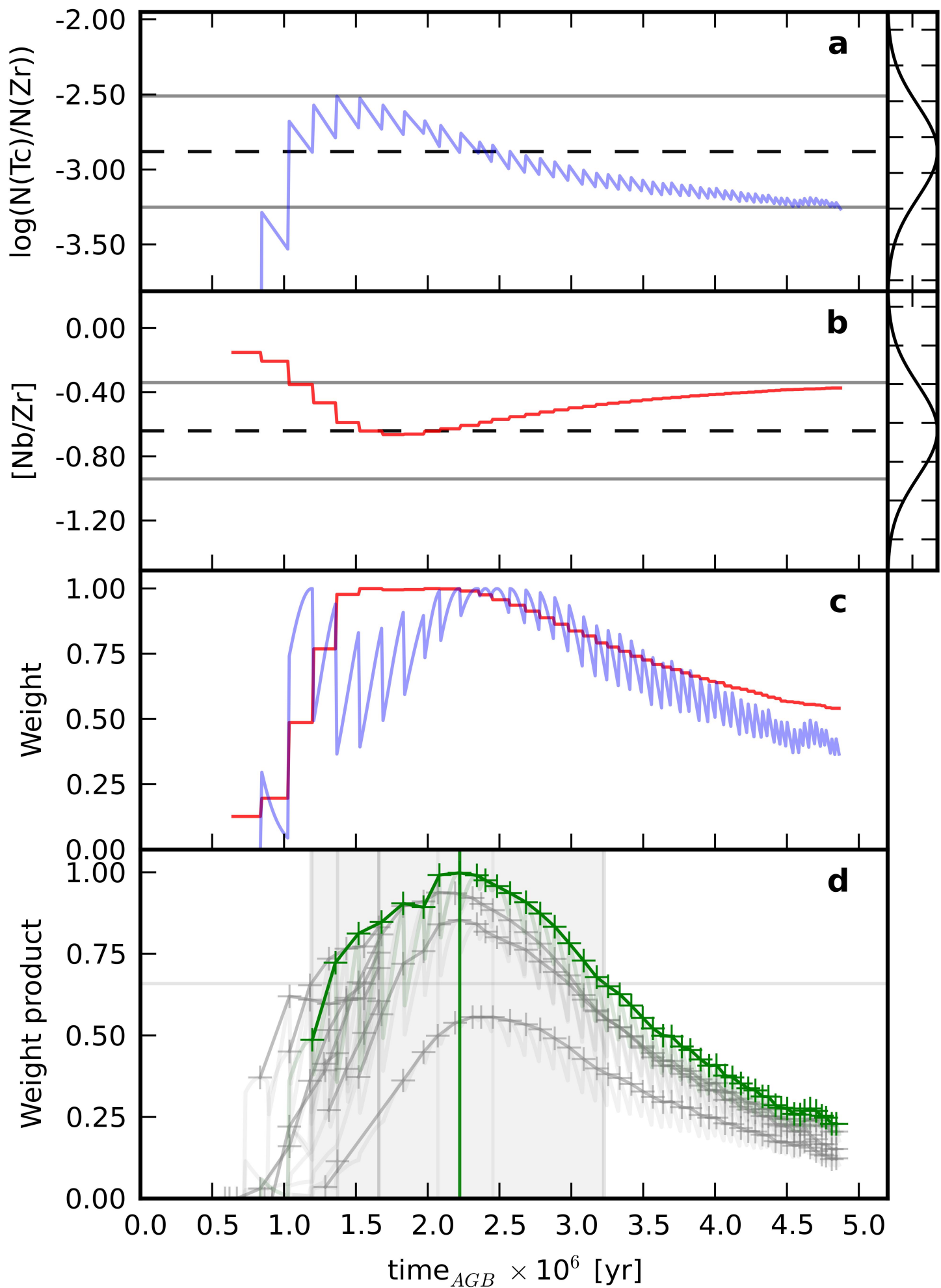